# In the Line of Fire: A Systematic Review and Meta-Analysis of Job Burnout Among Nurses

**Zahra Ghasemi Kooktapeh**

Management Faculty, Kharazmi University, Tehran, Iran.

Zahra.ghasemi0304@gmail.com

**Hakimeh Dustmohammadloo**

Department of Management, Unikl University Kuala Lumpur, Malaysia.

**Hooman Mehrdoost**

Department of Mechanical Engineering, Faculty of Engineering, Islamic Azad University, Mashhad, Iran.

**Farivar Fatehi**

Sheldon B. Lubar College of Business, University of Wisconsin Milwaukee, Milwaukee, WI, USA

Ffatehi@uwm.edu

**Abstract**

Using a systematic review and meta-analysis, this study investigates the impact of the COVID-19 pandemic on job burnout among nurses. We review healthcare articles following the PRISMA 2020 guidelines and identify the main aspects and factors of burnout among nurses during the pandemic. Using the Maslach Burnout questionnaire, we searched PubMed, ScienceDirect, and Google Scholar, three open-access databases, for relevant sources measuring emotional burnout, personal failure, and nurse depersonalization. Two reviewers extract and screen data from the sources and evaluate the risk of bias. The analysis reveals that 2.75% of nurses experienced job burnout during the pandemic, with a 95% confidence interval and rates varying from 1.87% to 7.75%. These findings emphasize the need for interventions to address the pandemic's effect on job burnout among nurses and enhance their well-being and healthcare quality. We recommend considering individual, organizational, and contextual factors influencing healthcare workers' burnout. Future research should focus on identifying effective interventions to lower burnout in nurses and other healthcare professionals during pandemics and high-stress situations.

## Introduction

Burnout syndrome poses a significant global challenge in human resource management (Rusca et al., 2019). It arises from prolonged exposure to work-related stress and is characterized by a decline in performance and functionality due to overwhelming psychological demands within the workplace (Panagioti et al., 2018). Bai et al. (2023) mention in their study that several factors underpin workplace pressures, with notable stressors including the challenges of "balancing work and family," "weak internal communications," "working hours/workload," and "weak leadership." Notably, healthcare service providers, including doctors, nurses, and medical staff, are particularly susceptible to burnout (Friganoviü et al., 2019; Jahanshahi et al., 2020; Chen et al., 2021; Zhang et al., 2020). Nurses, in particular, play a critical role on the front lines in combating illnesses (Zare et al., 2021), and they experience the highest incidence of burnout (Woo et al., 2020; Khaksar et al., 2010a), estimated at a staggering 54% probability (Rezaei et al., 2018a). The burnout syndrome encompasses a constellation of physical, mental, and emotional disorders that manifest as pessimism and a decline in the performance of healthcare professionals (Gómez Urquiza et al., 2017). Importantly, this deterioration in performance has far-reaching consequences, adversely affecting the quality of patient care and contributing to increased job turnover (Shajiei et al., 2020;

Hailay et al., 2020) due to less organizational commitment. According to Bai and Vahedian (2023), the relationship between organizational commitment and turnover is inverse. Employees who are more committed are less likely to leave the organization than those who are not. Their finding shows how important it is to increase organizational commitment among nurses to reduce turnover intentions.

The current global issue of nurse shortages is of significant concern, primarily driven by high attrition rates (Liu et al., 2018). This problem has been further exacerbated in the context of the COVID-19 pandemic (KILIÇ et al., 2021), with a distressing 31.9% mortality rate among nurses as of September 16, 20 (Galanis et al., 2020). Consequently, this situation presents a grave health policy crisis (Azari et al., 2021). Hence, healthcare system managers must prioritize the issue of burnout syndrome for two fundamental reasons. First, uphold the patients' rights charter and ensure high-quality services while safeguarding patients' rights (Gilavandi et al., 2019). Secondly, it is essential to support the diverse skills of healthcare professionals, particularly nurses, given the many responsibilities they shoulder within medical facilities (Gilavandi et al., 2019).

The advent of the COVID-19 pandemic has impacted nurses' attention, understanding, and decision-making abilities and has had chronic and ongoing repercussions on their physical, mental, and social well-being (Kang et al., 2020). Consequently, this has the potential to directly and indirectly diminish the quality and productivity of work environments, leading to a higher prevalence of burnout and attrition among nurses (Kang et al., 2020). Therefore, healthcare managers must identify the most important dimensions of job burnout to address this area effectively (Zare et al., 2021) and implement strategies to prevent and reduce this negative organizational phenomenon.

The causes and prevalence of burnout among nurses and medical staff have been widely investigated (López-López et al., 2019; Zeng et al., 2020; Khammar et al., 2018; Gómez-Urquiza et al., 2017; Hailay et al., 2020; Membrive-Jiménez et al., 2020; and Martos et al., 2020), but few studies have examined the factors of this organizational phenomenon in non-crisis situations in healthcare settings, as well as how the COVID-19 pandemic affects the prevalence, consequences, and symptoms of burnout among nurses. Therefore, identifying the most important dimensions of job burnout during the COVID-19 pandemic through a systematic review seems necessary.

## Literature Review

Job burnout is a discount in someone's potential to cope with traumatic factors, which manifests as physical and emotional exhaustion, terrible self-photo, bad attitude in the direction of work, and decreased social interplay with different employees (Huang et al., 2022; Jahanshahi et al., 2019; Dehkordy et al., 2013; Nawaser et al., 2015; Etemadi et al., 2022). Some of the signs and symptoms of task burnout syndrome consist of a lack of attention, difficulty in preserving statistics, common headaches, insomnia, fatigue, helplessness, low motivation, and self-doubt (Shoaib et al., 2017; Hakkak et al., 2016; Khaksar et al., 2010b; Moezzi et al., 2012). It has been broadly verified that burnout is a giant hassle for healthcare experts, especially nurses (Rana & Soodan, 2019; Gong et al., 2021). Consequently, nurses may showcase maladaptive behaviors due to activity burnout, resulting in extreme or persistent psychological strain with lengthy-term results (Li et al., 2018) and bad influences on their circle of relatives and social, non-public, and organizational lives.

Some of the essential outcomes of process burnout amongst nurses are absenteeism, turnover, common delays, various mental court cases, conflicts, task dissatisfaction, reduced pleasant of patient care, and interpersonal troubles with colleagues (Rudman et al., 2020; Gheitarani et al., 2023; Sepahvand et al., 2015; Daneshmandi et al., 2023; Vesal et al., 2013). Studies have suggested a 33.3% incidence of burnout among nurses, from 30 to 80% relying on the health center branch (Saeidi et al., 2020). Nurses who work in rotating shifts also experience higher tiers of burnout. Work-associated stressors, which include coping with unhealthy work conditions and interacting with sufferers, are among the challenges that nurses face (Jiang et al., 2017). Studies on burnout related to the COVID-19 pandemic among healthcare people have proven a tremendous and extensive relationship between pressure and burnout (Morgantini et al., 2020; Gheitarani et al., 2022). A study on the association between burnout, tension, and pressure problems during the COVID-19 pandemic showed that medical doctors and nurses enjoy excessive mental problems, including burnout (Sung et al., 2020). Bradley and Chahar (2020) emphasized the importance of the intellectual health of healthcare professionals at some stage in a pandemic to increase productivity and decrease burnout due to stress and uncertainty. However, because of the unprecedented nature of the COVID-19 pandemic, the extent and factors of burnout among nurses have now not been understood, especially concerning specific factors of labor environment and psychological responses (Wan et al., 2022; Shamsaddini et al., 2015; Sepahvand et al., 2023).

## Research Methodology

This study followed the systematic review and meta-analysis method based on the PRISMA2020 guidelines for review articles in healthcare. The search strategy consisted of the following steps:

1. Three databases were searched: PubMed, ScienceDirect, and Google Scholar.

2. The keywords were prevalence, burnout, nurses, and Maslach.

3. The search period covered the duration of the COVID-19 pandemic, from December 2019 to November 2021.

4. The initial search included reports on the prevalence of job burnout among nurses during the pandemic outbreak in both English and Farsi languages (indexed in the relevant databases). In the second step, only the open-access studies that used the Maslach Burnout Questionnaire (which measures burnout in three dimensions: emotional exhaustion, personal failure, and depersonalization) were selected for inclusion. The MBI questionnaire was chosen because it is the most common for assessing the prevalence of job burnout and has good validity and reliability (Pisanti et al., 2013). It is important to note that using other questionnaires to assess job burnout would introduce more uncertainty in the synthesis and analysis of results (Barbara et al., 2014). The studies that did not provide sufficient quantitative information in the text or tables were also excluded from the systematic review and meta-analysis.

5. To make the final selection, two team members independently and concurrently searched and reviewed the articles. Articles were selected for inclusion based on three stages: title, abstract, and full-text review; in case of disagreement, an arbitrator resolved it. The steps of removing articles can be seen in Figure 1.

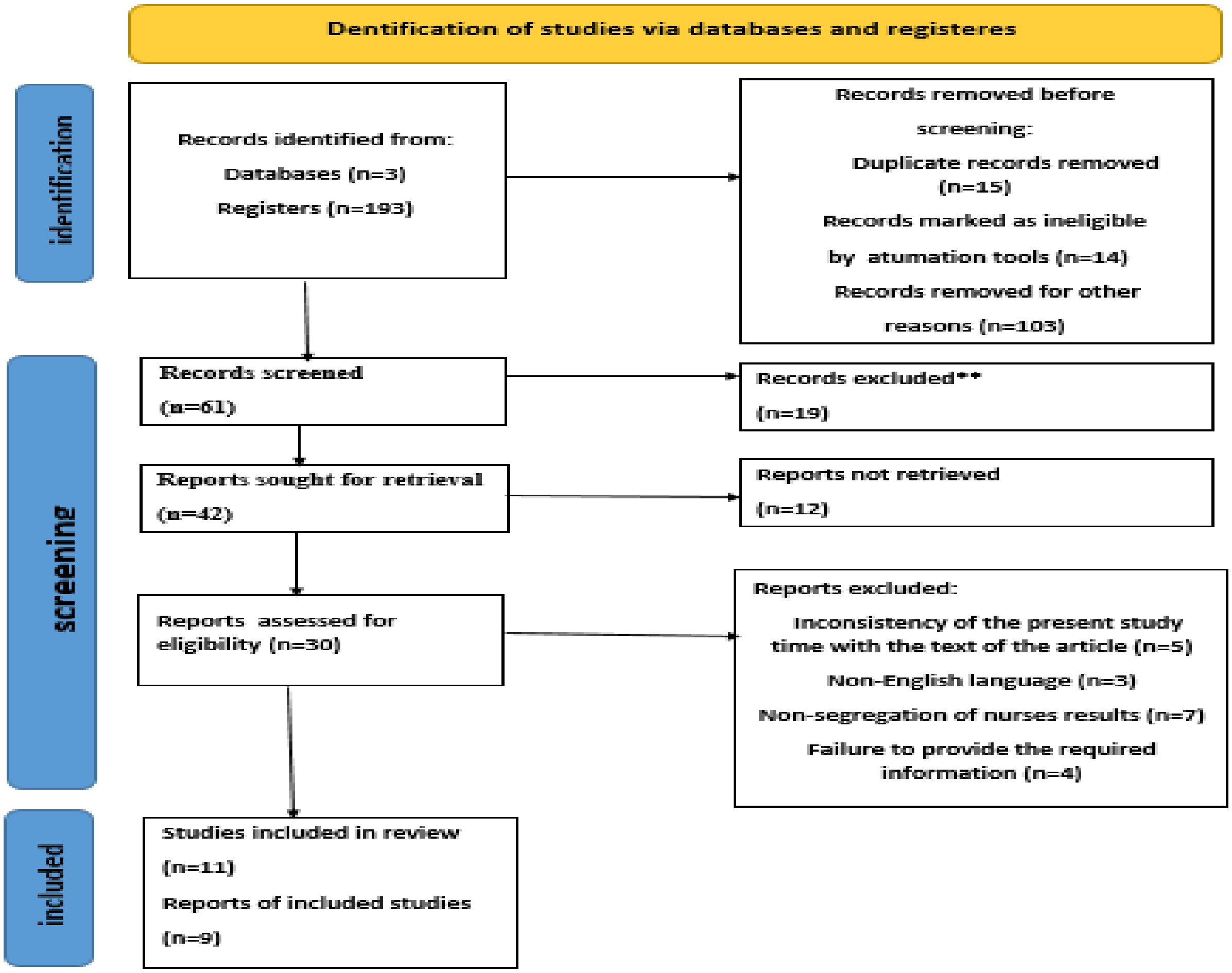

**Figure 1:** The flowchart of the preferred items in the report of systematic review and meta-analysis of articles.

## Quality Evaluation

Out of the 193 articles initially retrieved, 61 articles were duplicates and were removed. In the second screening stage, 31 studies were excluded from the review process for various reasons. In the third screening stage, 19 studies were excluded due to a mismatch of the study period, use of non-English languages, and lack of separation between the results of nurses and other employees. Finally, 14 articles that met the inclusion criteria were selected for the meta-analysis. All the final studies, except for one from China, were from developing countries. All studies used Maslach's job burnout questionnaire to measure the dimensions of job burnout. Nine articles were ultimately included in our study.

## Findings

As shown in Table 1, the prevalence of burnout among nurses ranged from 57.7% to 87.1%, with a 95% confidence interval ($p = 0.007$). The overall prevalence of burnout was 75.2%. The results indicated that all studies except for study number 2 ($p = 0.462$) were statistically significant and relevant.

**Table 1:** Prevalence, significance, and standard error of the nine studies included in the analysis.

| Model | Study name | Cumulative statistics | | | | |
|---|---|---|---|---|---|---|
| | | Point | Lower limit | Upper limit | Z-Value | p-Value |
| | 1.000 | 0.683 | 0.658 | 0.708 | 12.855 | 0.000 |
| | 2.000 | 0.584 | 0.363 | 0.776 | 0.736 | 0.462 |
| | 3.000 | 0.754 | 0.502 | 0.903 | 1.973 | 0.049 |
| | 4.000 | 0.881 | 0.666 | 0.965 | 2.997 | 0.003 |
| | 5.000 | 0.931 | 0.781 | 0.981 | 3.820 | 0.000 |
| | 6.000 | 0.940 | 0.815 | 0.983 | 4.234 | 0.000 |
| | 7.000 | 0.813 | 0.708 | 0.886 | 4.926 | 0.000 |
| | 8.000 | 0.797 | 0.617 | 0.906 | 3.003 | 0.003 |
| | 9.000 | 0.752 | 0.577 | 0.871 | 2.717 | 0.007 |
| Random | | 0.752 | 0.577 | 0.871 | 2.717 | 0.007 |

Cumulative event rate (95% CI): -1.00 -0.50 0.00 0.50 1.00

In chart number 2, the prevalence of job burnout among nurses has been investigated in the selected nine studies.

**Figure 2:** The prevalence of job burnout among nurses in the selected nine studies.

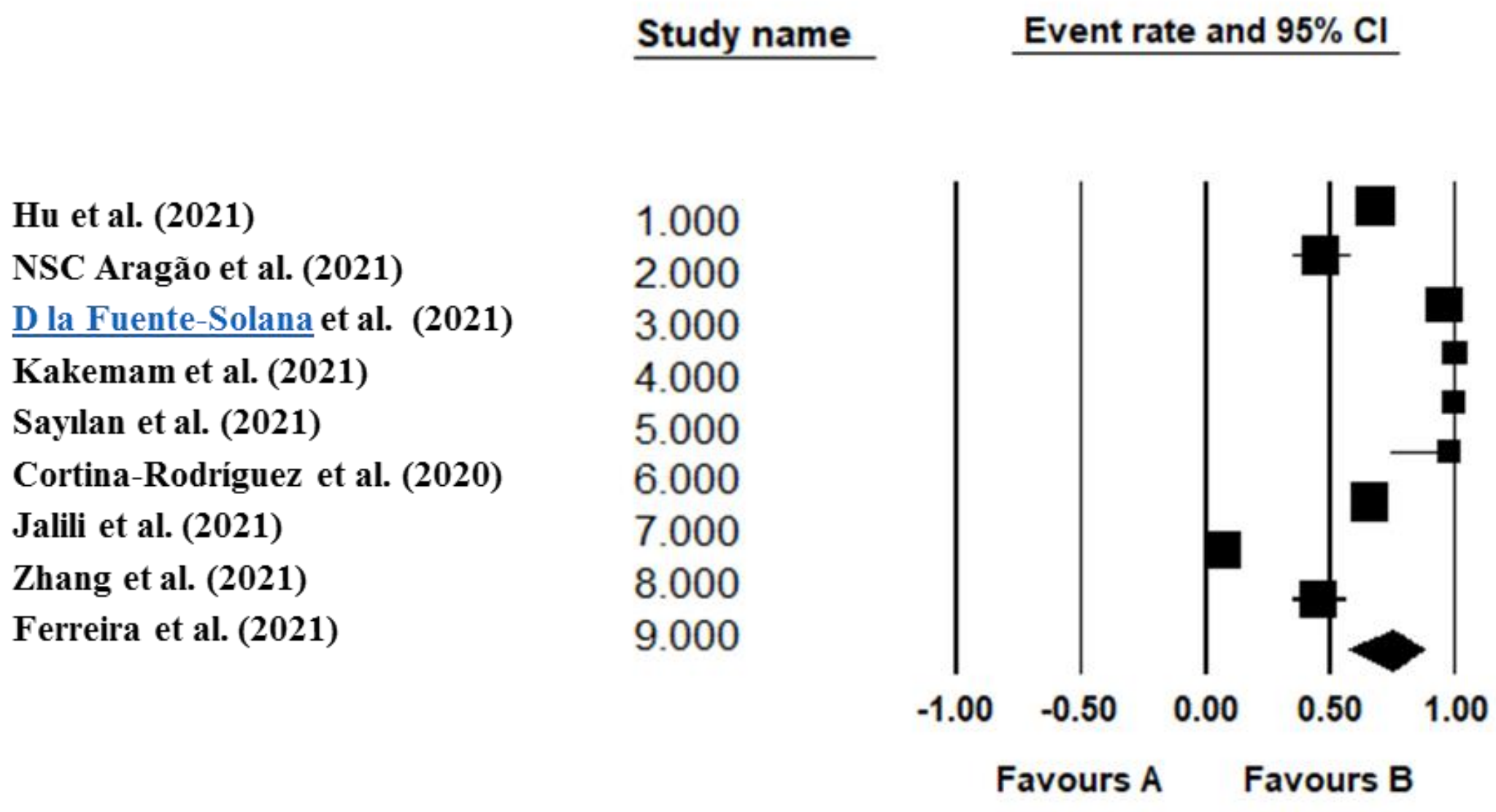

According to Table 1 and Figure 2, the line segments show the confidence interval of the prevalence of burnout in each study; the middle point of each line segment estimates the prevalence of job burnout in each study; and the diamond symbol shows the prevalence of job burnout for the entire study. The lowest prevalence of occupational depression was in the second study (NSC Aragão et al., 2021); the highest prevalence of job burnout among nurses was also in the sixth study (Cortina-Rodríguez et al., 2020). The total prevalence was 75.2%, which is a significant number. The seventh study related to Iran had low and high prevalence rates of burnout of about 0.708 and 0.886, respectively.

Figure 3 illustrates the funnel plot related to the prevalence of job burnout in the selected nine studies based on the Egger Regression test. As seen, the symmetrical funnel plot indicates the absence of publication bias, and therefore, no evidence of publication bias was observed in the present study ($p > 0.05$ for Egger's weighted regression analysis).

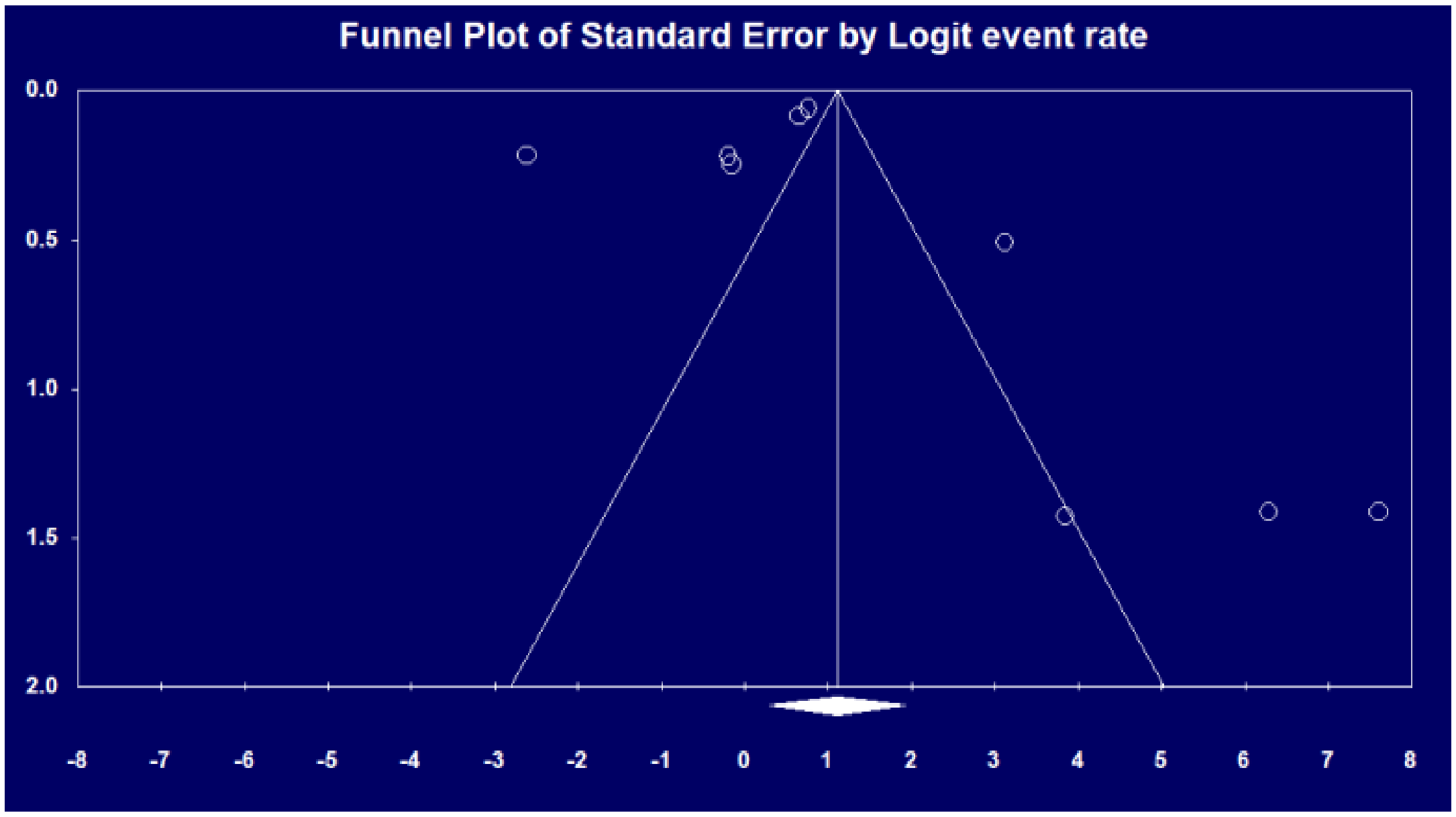

**Figure 3:** Funnel chart related to the spread of job burnout prevalence in the nine studies based on Egger's regression test.

## Conclusion

Systematic review and meta-analysis can facilitate changes in clinical practice, knowledge advancement, and better management in the field of health and disease. The main objective of this study was to enhance the knowledge of the prevalence of job burnout among nurses during the COVID-19 pandemic. The results revealed that the nurses experienced high levels of job burnout due to the pandemic. Several factors, such as social, cultural, and occupational aspects, influenced the burnout levels. The COVID-19 pandemic has been a significant challenge for nurses and healthcare managers worldwide, and learning from the experiences gained during the multiple waves of the pandemic in different countries can be essential to devise better strategies for future waves.

Reducing burnout among nurses through implementing interventions may be an effective strategy to improve the quality of patient care and lower burnout rates, especially in public health centers (Staines et al., 2020). In this regard, nursing managers are urgently required to create a supportive work environment that fosters the professional development of nurses. Primary approaches are needed to minimize burnout, such as access to psychosocial support, web-based services, psychological first aid, psychological support hotlines, and self-care techniques. Quality improvement and patient safety teams can also help healthcare workers cope with the pandemic (Gurses et al., 2020). Therefore, several measures are recommended, such as mental health screening and early support interventions for high-risk nurses, immediate access to health care services, designated rest periods, social support through hospital support groups to reduce depression, and adequate personal protective equipment for all. Since successive waves of the pandemic are likely to occur, it is necessary to restore health professionals' mental health to prevent the collapse of health systems. Governments, healthcare organizations, and policymakers should control the current situation and prepare healthcare systems, workers, and nurses to better respond to the COVID-19 pandemic.